\documentclass[aps,prc,fleqn,floatfix,twocolumn,superscriptaddress,twoside]{revtex4}

\usepackage[]{graphicx}% Include figure files

\usepackage{amsmath,amssymb,amsfonts}

\usepackage{color}

\begin{document}

\title{Jet transport and photon bremsstrahlung via longitudinal and transverse scattering}

\author{Guang-You Qin}

\affiliation{Institute of Particle Physics and Key Laboratory of Quark and Lepton Physics (MOE), Central China Normal University, Wuhan, 430079, China  }
\affiliation{Department of Physics and Astronomy, Wayne State University, Detroit, MI, 48201.}

\author{Abhijit Majumder}

\affiliation{Department of Physics and Astronomy, Wayne State University, Detroit, MI, 48201.}

\date{\today}
%%%%%%%%%%%%%%%%%%%%%%%%%%%%%%%%%%%%%%%%%%%%%%%%%%%%%%%%%%%%%%%%%%%%%%%%%%%%%%%%
%%%%%%%%%%%%%%%%%%%%%%%%%%%%%%%%%%%%%%%%
\begin{abstract}

We study the effect of multiple scatterings on the propagation of hard partons and the production of jet-bremsstrahlung photons inside a dense medium in the framework of deep-inelastic scattering off a large nucleus.
We include the momentum exchanges in both longitudinal and transverse directions between the hard partons and the constituents of the medium.
Keeping up to the second order in a momentum gradient expansion, we derive the spectrum for the photon emission from a hard quark jet when traversing dense nuclear matter.
Our calculation demonstrates that the photon bremsstrahlung process is influenced not only by the transverse momentum diffusion of the propagating hard parton, but also by the longitudinal drag and diffusion of the parton momentum.
A notable outcome is that the longitudinal drag tends to reduce the amount of stimulated emission from the hard parton.

\end{abstract}
\maketitle
%%%%%%%%%%%%%%%%%%%%%%%%%%%%%%%%%%%%%%%%%%%%%%%%%%%%%%%%%%%%%%%%%%%%%%%%%%%%%%%%
%%%%%%%%%%%%%%%%%%%%%%%%%%%%%%%%%%%%%%%%

\section{Introduction}

The modification of hard partonic jets in dense media provides a useful tool for studying the properties of the highly excited nuclear matter produced in relativistic heavy-ion collisions.
One of the primary experimental signatures for jet modification is the significant depletion of the high transverse momentum ($p_T$) hadron yield in heavy-ion collisions compared to that in binary-scaled proton-proton collisions \cite{Adcox:2001jp,Adler:2002xw,Aamodt:2010jd}.
Such depletion has been attributed to the loss of the forward momentum and energy by the hard partons in the dense nuclear medium they traversed before fragmenting into hadrons \cite{Gyulassy:1990ye,Wang:1991xy}.

By all accounts, one expects the energy loss to originate from a combination of elastic collisions \cite{Bjorken:1982tu, Braaten:1991we, Djordjevic:2006tw, Qin:2007rn, Schenke:2009ik} and inelastic radiative processes \cite{Gyulassy:1990ye,Wang:1991xy, Baier:1996kr, Baier:1996sk, Zakharov:1996fv}, that hard jets experience when propagating through the dense media.
For light partons, medium-induced gluon radiation is thought to be the more dominant mechanism for parton energy loss.
For heavy flavor partons with large finite masses, collisional energy loss appears to be equally or even more important than medium-induced gluon radiation, especially in the low and intermediate $p_T$ regime \cite{Moore:2004tg, Mustafa:2004dr, Wicks:2005gt, Qin:2009gw, Cao:2013ita}.

There are currently a few different schemes for the study of medium-induced gluon bremsstrahlung (the radiative part of parton energy loss) in a dense nuclear medium: the Baier-Dokshitzer-Mueller-Peigne-Schiff-Zakharov (BDMPS-Z) approach~\cite{Baier:1996kr, Baier:1996sk, Zakharov:1996fv}, the Gyulassy, Levai and Vitev (GLV) approach~\cite{Gyulassy:1999zd, Gyulassy:2000fs, Gyulassy:2000er}, the Amesto-Salgado-Wiedemann (ASW) approach~\cite{Wiedemann:2000za,Wiedemann:2000tf}, the Arnold-Moore-Yaffe (AMY) approach~\cite{Arnold:2001ba, Arnold:2002ja} and the higher twist (HT) approach~\cite{Guo:2000nz, Wang:2001ifa, Majumder:2009ge}.
Detailed comparisons of the different formalisms may be found in Ref. \cite{Armesto:2011ht}.
Sophisticated phenomenological  analyses have been performed for various jet quenching observables, such as single inclusive hadron suppression \cite{Bass:2008rv, Armesto:2009zi, Chen:2010te}, as well as dihadron and photon-hadron correlations \cite{Zhang:2007ja, Majumder:2004pt, Qin:2009bk, Renk:2008xq}.
One of the main goals of these studies is to quantitatively extract the jet transport coefficients, such as $\hat{q}$ and $\hat{e}$, in dense media by comparing with the data on jet modification.
This requires that the originating formalisms contain a representative description of all physical processes that influence the partonic substructure of the jet.

While different parton energy loss models mentioned above have rather different origins and make slightly different approximations, the calculations of radiative energy loss have so far only focused on the gluon radiation induced by transverse momentum diffusion, experienced by the propagating hard partons in the dense medium.
In fact, when a jet scatters off the medium constituents, it is not only the transverse momentum, but also the longitudinal momentum that are exchanged between the jet and the medium \cite{Qin:2012fua, Majumder:2008zg, Abir:2014sxa}.
In many calculations, the transfer of longitudinal momentum has only been considered in the evaluation of purely collisional energy loss.
There have been studies on the longitudinal momentum loss experienced by radiated partons in the context of jet shower evolution \cite{Qin:2009uh, Neufeld:2009ep, Qin:2010mn, Qin:2012gp}.
However, the influence of exchanging longitudinal momentum on the stimulated radiation vertex has not yet been throughly considered.
In this work, we re-investigate the medium-induced radiation from a light parton, by taking into account the influence from both transverse and longitudinal momentum transfers, when the parton undergoes multiple scattering with the medium, in the radiation process.

As a step up to the more complicated calculation of gluon radiation, we study the problem of real photon radiation from a parton which propagates through a dense extended medium and experiences multiple scatterings off the glue field of the medium. While the emitted photon escapes the medium without further strong-interaction with medium, very unlike the case of a radiated gluon, such a problem does encode many interesting features of the medium-induced gluon radiation, such as the Landau-Pomeranchuck-Migidal (LPM) effect, as well as the momentum approximation schemes involved in the problem, which are the same as the problem with gluon radiation. Therefore, the photon calculation serves as an intermediate step for the further investigation of medium-induced gluon emission from a hard jet. In this regard, we will explore the effect of a larger than usual longitudinal exchange with the medium: In most treatments, the longitudinal momentum exchanged with the medium is considerably smaller than the transverse momentum exchanged (Glauber-Coulomb scattering), however, in this paper, we will allow one component of the longitudinal momentum to be of the order of the transverse momentum exchanged.

In addition, photon bremsstrahlung from a high $p_T$ jet is itself of great interest.
Such process has shown to be an important source for the photon production in relativistic heavy-ion collisions, especially in the intermediate $p_T$ regime \cite{Fries:2002kt, Qin:2009bk}.
As the production of such photons involves the interaction between hard jets and the dense medium, it also opens an electromagnetic window to probe the jet transport properties in dense media.
It is also of great relevance to the study of photon-jet correlations \cite{Renk:2006qg, Zhang:2009rn, Qin:2009bk}, which are expected to provide more stringent characterizations of the quenching experienced by the jet in the dense medium, than single hadron or dihadron observables.

The paper is organized as follows. In Sec. II the calculation of photon bremsstrahlung at leading-twist is presented in some details. In Sec. III, photon radiation from a hard parton which experiences a number of scatterings in the dense medium is analyzed. In Sec. IV, the sum of photon production points is carried out. In Sec. V, Taylor expansion for the hard part is invoked and the number of multiple scattering is resummed to obtain the final expression for the photon bremsstrahlung spectrum from a hard jet in dense medium. Sec. VI contains our conclusion and discussions.

\section{Photon bremsstrahlung at leading-twist}

In this section, we provide some details for the calculation of photon bremsstrahlung from a hard parton produced in the deep-inelastic scattering (DIS) off a large nucleus at leading twist.
Consider the semi-inclusive process of the DIS off a nucleus in which one hard quark with a momentum $l_q$ and a bremsstrahlung photon with momentum $l$ are produced,
\begin{eqnarray}
e(L_1) + A(P_A) \to e(L_2) + q(l_q) + \gamma(l) + X.
\end{eqnarray}
Here $L_1$ and $L_2$ denote the momenta of the incoming and outgoing leptons.
The incoming nucleus with atomic number $A$ has a momentum $P_A=Ap$, and each nucleus inside nucleus $A$ carries a momentum $p$.

Throughout this work, we utilize the light-cone component notation for four vectors ($p = [p^+, p^-, \mathbf{p}_\perp]$), where
\begin{eqnarray}
p^+ = (E+p_z)/\sqrt{2}, p^- = (E-p_z)/\sqrt{2}.
\end{eqnarray}
In the Briet frame of the DIS, the incoming virtual photon $\gamma^*$ has a momentum $q$ and the nucleus has a momentum $P_A$:
\begin{eqnarray}
&& q = L_2 - L_1 = [-x_Bp^+, q^-, \mathbf{0}_\perp], \nonumber \\
&& P_A = A[p^+, 0, \mathbf{0}],
\end{eqnarray}
where $x_B$ is the Bjorken variable, $x_B = Q^2 / (2p^+q^-)$, with $Q^2$ the invariant mass of the virtual photon.
The radiated photon has a momentum $l$ and carries a fraction $y$ of the quark forward momentum $q^-$, i.e., $y=l^-/q^-$.

The double differential cross section of such semi-inclusive DIS process may be expressed as
\begin{eqnarray}
\frac{E_{L_2} d\sigma}{d^3L_2 d^3 l_q d^3 l} = \frac{\alpha_{e}}{2\pi s} \frac{1}{Q^4} L_{\mu\nu} \frac{dW^{\mu\nu}}{d^3l_q d^3l},
\end{eqnarray}
where $s=(p+L_1)^2$ is the total invariant mass of the lepton-nucleon collision system, and $\alpha_e$ is the electromagnetic coupling. The leptonic tensor is give by
\begin{eqnarray}
L_{\mu\nu} = \frac{1} {2}{\rm Tr}[\gamma \cdot L_1 \gamma_\mu \gamma \cdot L_2 \gamma_\nu].
\end{eqnarray}
The semi-inclusive hadronic tensor is defined as
\begin{eqnarray}
W^{\mu\nu} \!\!&=&\!\! \sum_{X} (2\pi)^4 \delta^4 (q + P_A - P_X) \nonumber\\ \!\!&\times&\!\!
\langle A| J^\mu(0)|X\rangle \langle X|J^\nu(0)|A\rangle.
\end{eqnarray}
Here $|A\rangle$ represent the initial state of an incoming nucleus $A$, averaged over spins.
$|X\rangle$ represents the general final hadronic or partonic states, with $\sum_X$ running over all possible final states except the stuck hard jet and the emitted real photon.
$J^\mu = Q_q \psi_q \gamma^\mu \psi$ is the hadron electromagnetic current, with $Q_q$ the charge of a quark of flavor $q$ in units of the positron charge $e$.
Since we are interested in the final state interaction between the medium and the stuck quark, our focus will be on the hadronic tensor.
In the following, the leptonic tensor will not be discussed further.

\begin{figure}[thb]
\includegraphics[width=1.0\linewidth]{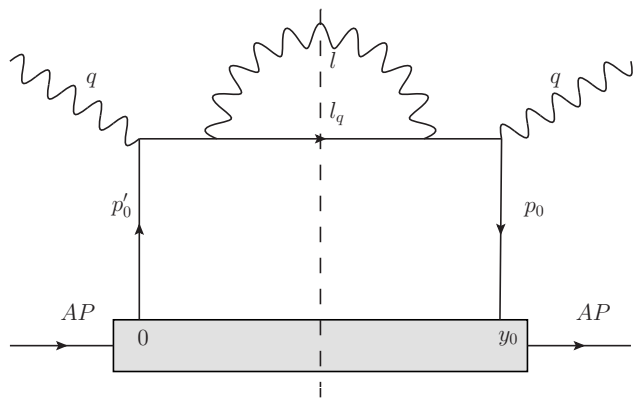}
 \caption{Leading twist contribution to the semi-inclusive DIS hadronic tensor $W^{\mu\nu}$.
} \label{leadingtwist}
\end{figure}

Fig. \ref{leadingtwist} shows the leading-twist contribution to the photon bremsstrahlung in the semi-inclusive DIS.
It represents the process where a hard quark, produced from one nucleon of the nucleus, radiates a hard photon and then exits the nucleus without further interaction.
In this study, we choose the light cone gauge, and other types of diagrams are suppressed.
The leading-twist contribution to the hadronic tensor may be expressed as:
\begin{eqnarray}
W_0^{A \mu\nu}  \!\!&=&\!\!  \sum_q Q_q^4 e^2 \int \frac{d^4l}{(2\pi)^4} (2\pi)\delta(l^2) \int \frac{d^4l_q}{(2\pi)^4} (2\pi) \delta(l_q^2) \nonumber\\\!\!&\times&\!\!
\int d^4y_0 e^{iq\cdot y_0}  \int d^4z \int d^4z' \int \frac{d^4 q_1}{(2\pi)^4} \int \frac{d^4 q_1'}{(2\pi)^4} \nonumber\\\!\!&\times&\!\!
e^{-i q_1 \cdot (y_0-z)} e^{-i q_1' \cdot (z'-y_0')} e^{-il_q\cdot (z-z')} e^{-i l\cdot (z-z')}  \nonumber\\\!\!&\times&\!\!
\langle A |\bar{\psi}(y_0) \gamma^\mu \frac{\gamma \cdot q_1}{q_1^2 - i\epsilon} \gamma^\alpha (\gamma \cdot l_q) \gamma^\beta \frac{\gamma \cdot q_1'}{q_1^{'2} - i\epsilon} \gamma^\nu \psi(0)|A\rangle \nonumber\\\!\!&\times&\!\!
G_{\alpha\beta}(l).
\end{eqnarray}
Here the factor $G_{\alpha\beta}(l)$ represents the sum of the radiated photon's polarizations.
In the light cone gauge ($A^-=0$), it is given by
\begin{eqnarray}
G_{\alpha\beta}(l) = -g_{\alpha \beta} + \frac{n_\alpha l_\beta + n_\beta l_\alpha}{n\cdot l},
\end{eqnarray}
where we have defined the light cone vector $n=[1,0,\mathbf{0}]$, thus we have $n\cdot l = l^-$.
With such setup, the largest component of the vector potentials from the initial states is $A^+$ component.

We may perform the integrations over the photon radiation positions $z$, $z'$ which give two $\delta$ functions.
They can be utilized to perform the integrations over the momenta $q_1$, $q_1'$ which set $q_1 = q_1' = p_0 + q$.
Then the hadronic tensor reads,
\begin{eqnarray}
W_0^{A \mu\nu} \!\!&=&\!\! \sum_q Q_q^4 e^2 \int \frac{d^4l}{(2\pi)^4} (2\pi)\delta(l^2) \int \frac{d^4l_q}{(2\pi)^4} (2\pi) \delta(l_q^2)\nonumber\\\!\!&\times&\!\!
\int \frac{d^4p_0}{(2\pi)^4} (2\pi)^4 \delta^4(q + p_0 - l_q -l) \int d^4y_0 e^{-ip_0\cdot y_0} \nonumber\\\!\!&\times&\!\!
\langle A |\bar{\psi}(y_0) \gamma^\mu \frac{\gamma \cdot q_1}{q_1^2 - i\epsilon} \gamma^\alpha (\gamma \cdot l_q) \gamma^\beta \frac{\gamma \cdot q_1'}{q_1^{'2} - i\epsilon} \gamma^\nu \psi(0)|A\rangle \hspace{-12pt}\nonumber\\\!\!&\times&\!\!
G_{\alpha\beta}(l).
\end{eqnarray}
Here we have re-introduced the variable $p_0$ from the following identity,
\begin{eqnarray}
 \int \frac{d^4p_0}{(2\pi)^4} (2\pi)^4 \delta^4(q + p_0 - l_q -l) = 1.
\end{eqnarray}
The above four-$\delta$-function can then be used to perform the integration over $l_q$.
The on-shell condition for the photon, $\delta(l^2)$, can be used to do the integration over $l^+$ which gives $l^+ = l_\perp^2 / (2l^-)$.
The on-shell $\delta(l_q^2)$ function for the final quark gives,
\begin{eqnarray}
(2\pi) \delta(l_q^2) = \frac{(2\pi)}{2 p^+q^-(1-y)} \delta(x_0 - x_B - x_L).
\end{eqnarray}
Here we have introduced the new momentum fraction $x_L$, defined as $x_L = l_\perp^2/[2p^+q^-y(1-y)]$.
It is related to the photon formation time as $\tau_{\rm form} = 1/(x_Lp^+)$.
In the very high energy and collinear limit, the expression can be simplified via $p_0^+ = x_0 p^+$,  $p_0^- \to 0$ and $\mathbf{p}_{0\perp} \to 0$.
In this limit, one may ignore the $\perp$-component of quark field operators,
\begin{eqnarray}
\!\!&&\!\! \langle A| \bar{\psi}(y_0) \cdots \psi(0) |A\rangle = A C_p^A \langle p| \bar{\psi}(y_0) \cdots \psi(0) |p\rangle
\nonumber\\\!\!&&\!\!
\approx A C_p^A \langle p| \bar{\psi}(y_0) \frac{\gamma^+}{2} \psi(0) |p\rangle {\rm Tr}[\frac{\gamma^-}{2} \cdots],
\end{eqnarray}
where $\gamma^+$ and $\gamma^-$ are used to project the leading spin projection,
\begin{eqnarray}
\gamma^+ = (\gamma^0 + \gamma^3)/\sqrt{2}, \gamma^- = (\gamma^0 - \gamma^3)/\sqrt{2}.
\end{eqnarray}
The factor $C_p^A$ represents the probability to find a nucleon state with momentum $p$ inside a nucleus with $A$ nucleons.

The above $\delta$-function can be applied to perform the integration over $p_0^+ = p^+ x_0$.
We further perform the integrations over $y_0^+$ and $\mathbf{y}_{0\perp}$ and obtain a three-$\delta$-function,
which allows us to carry out the integration over $p_0^-$, $\mathbf{p}_{0\perp}$.
Now the hadronic tensor reads,
\begin{eqnarray}
W_0^{A \mu\nu} \!\!&=&\!\! A C_p^A \sum_q Q_q^4 \frac{\alpha_e}{2\pi} \int dy \int \frac{d^2l_\perp}{\pi l_\perp^2} \frac{(2\pi) f_q(x_B+x_L)}{8p^+(q^-)^2x_L} \nonumber\\\!\!&\times&\!\!
 {\rm Tr} [ \frac{\gamma^-}{2} \gamma^\mu (\gamma \cdot q_1) \gamma^\alpha (\gamma \cdot l_q) \gamma^\beta (\gamma \cdot q_1')\gamma^\nu ] G_{\alpha\beta}(l). \ \ \ \
\end{eqnarray}
Here $f_q(x)$ represents the parton distribution function of a quark with flavor $q$ in a nucleon, with a fraction $x$ of the forward momentum $p^+$ of the nucleon,
\begin{eqnarray}
f_q(x) = \int \frac{dy_0^-}{2\pi} e^{-ixp^+y_0^-} \langle p| \bar{\psi}(y_0^-) \frac{\gamma^+}{2} \psi(0)|p \rangle.
\end{eqnarray}
The trace combined with $G_{\alpha\beta}$ can be carried out by using the commutation relations of $\gamma$ matrices: $\{\gamma^+, \gamma^-\} = 2$, $\{\gamma^\pm, \gamma^\pm\} = 0$, and $\{\gamma^\pm, \gamma_\perp\} = 0$,
\begin{eqnarray}
\!\!&&\!\!  {\rm Tr} [ \frac{\gamma^-}{2} \gamma^\mu (\gamma \cdot q_1) \gamma^\alpha (\gamma \cdot l_q) \gamma^\beta (\gamma \cdot q_1')\gamma^\nu ] G_{\alpha\beta}(l)
\nonumber\\ &&  = (-g_\perp^{\mu\nu}) [8p^+(q^-)^2 x_L] P(y).
\end{eqnarray}
The factor $P(y)=[1+(1-y)^2]/y$ is the quark-to-photon splitting function, which represents the probability that a quark radiates a photon carrying away a faction of $y$ of its forward momentum.
The projection tensor $g_\perp^{\mu\nu}$ is defined as $g_\perp^{\mu\nu} = g^{\mu\nu} - g^{\mu-}g^{\nu+} - g^{\mu+} g^{\nu-}$.
The hadronic tensor is now obtained as,
\begin{eqnarray}
\frac{dW_0^{A\mu\nu}}{d^2l_\perp dy} = A C_p^A \sum_q Q_q^4  \frac{\alpha_e}{2\pi} \frac{P(y)}{\pi l_\perp^2} (-g_\perp^{\mu\nu}) (2\pi) f_q(x_B+x_L). \hspace{-12pt}\nonumber\\
\end{eqnarray}

Since the parton is produced immediately after the initial hard scattering, its momentum is related to photon's momentum by $l_q^- = q^-(1-y)$ and $\mathbf{l}_{q \perp} + \mathbf{l}_{\perp} = 0$. Therefore, the double-differential hadronic tensor for the momentum distribution of the final quark is given by
\begin{eqnarray}
\frac{dW_0^{A \mu\nu}}{d^2l_\perp dy d^2l_{q \perp} dl_q^-} = \frac{dW_0^{A \mu\nu}}{d^2l_\perp dy} \delta(l_q^- - q^-(1 - y)) \delta^2(\mathbf{l}_{q \perp} + \mathbf{l}_\perp).
\hspace{-12pt}\nonumber\\
\end{eqnarray}
In the following, we will investigate how the multiple scatterings from the medium change the momentum distribution of the final quark and the spectrum of photon radiation.

\section{Photon bremsstrahlung from multiple scatterings}

In this section, we derive the single photon radiation spectrum from a hard jet which undergo multiple scatterings from the dense medium.
Higher twist contribution is obtained from those diagrams which contain the products with more partonic operators in the medium.
Usually, such contributions are suppressed by powers of the hard scale $Q^2$, but in an extended medium, a sub-class of these contributions may be enhanced by the longitudinal extent traveled by the stuck quark.
Our focus in this work will be to isolate and resum such length-enhanced higher twist contribution.

The generic diagram being considered here is shown in Fig.~\ref{mntwist}. It describes the process that a hard virtual photon strikes a quark in the nucleus
with momentum $p_0'$ ($p_0$ in complex conjugate) at location $y_0'=0$ ($y_0$ in complex conjugate).
The stuck quark is sent back to the nucleus, carrying momentum $q_1'$ ($q_1$ in the complex conjugate).
During its propagation through the nucleus, the hard quark scatters off the gluon field at locations $y_j'$ with $0<j<m$ ($y_i$ in the complex conjugate with $0<i<n$).
Through each scattering, the hard quark picks up momentum $p_j'$ ($p_i$ in the complex conjugate).
The photon is radiated between locations $q$ and $q+1$ in the amplitude ($p$ and $p+1$ in the complex conjugate).

In this section, and through most of the next, we will retain terms at all orders and with no approximation as to the momentum structure of the exchanged momenta. 
At the end of Sect.~\ref{sum-sect} and in Sect.~\ref{resum-sect}, we will approximate and attempt to extract a closed form expression. Here we will invoke the 
power counting of the exchanged gluons in Fig~\ref{mntwist}, which lie in the Glauber limit~\cite{Idilbi:2008vm}. 
In so doing we will use $Q$ to refer to the hardest scale in the problem and $\lambda$ to refer to a small dimensionless parameter, which in product with $Q$ will refer to a softer scale. 
In the case of a near on-shell projectile parton (with a large near on-shell momentum $q^{-} \sim Q$ and $q^{+} \sim \lambda^{2} Q$) scattering 
off another near on-shell target parton traveling in the opposite direction (with $p^{+} \sim Q$ and $p^{-} \sim \lambda^{2} Q$), 
the exchanged gluons have the standard Glauber scaling of momenta, 
\begin{eqnarray}
k^{+} \sim \lambda^{2} Q ; \,\,\, k^{-} \sim \lambda^{2} Q; \,\,\, k_{\perp} = \lambda Q.
\end{eqnarray}
If one were to drop the requirement that the target be an on shell parton, then a larger $(-)$-component of the exchanged gluon can be absorbed, leaving the projectile parton near on shell. This will be the scaling used in this paper, sometimes referred to as a \emph{longitudinal}-Glauber. The scaling for such gluons is given as,
\begin{eqnarray} 
k^{+} \sim \lambda^{2} Q ; \,\,\, k^{-} \sim \lambda Q; \,\,\, k_{\perp} = \lambda Q.\label{long-glauber}
\end{eqnarray}
It is this second scaling that is assumed in this paper. The goal is to test if a relaxation of the energy deposition rate in the medium can noticeably change the 
photon (and by extension the gluon) radiation spectrum. In what follows, the scaling of all exchanged momenta and gauge fields will be driven by the assumptions of 
Eq.~\eqref{long-glauber} for the exchanged gluon.

Following the notation of the previous section where the photon momentum is denoted as $l$ and the final quark's  momentum $l_q$, the momenta of the quark lines after each scattering at location $y'_j$ are denoted as: $q'_{j+1}$ before photon emission and $\bar{q}'_{j+1}$ after the photon emission ($y_i$, $q_{i+1}$, and $\bar{q}_{i+1}$ in the complex conjugate).
For the quark lines immediately around the photon radiation,
they are denoted as $p'_{q+1}$ and $\bar{p}'_{q+1}$ ($p_{p+1}$ and $\bar{p}_{p+1}$ in the complex conjugate),
where $\bar{q}_{p+1} = q_{p+1}-l$ and $\bar{q}'_{q+1} = q'_{q+1}-l$.
From the momentum conservation at each vertex, various momenta denoted in the picture have the following relations:
\begin{widetext}
\begin{eqnarray}
q_{i+1} = q + K_i = q + \sum_{j=0}^i p_j, \,\,\,\, \forall i \le p+1 ; \,\,\,\, & \bar{q}_{i+1} = q + K_i - l = q + \sum_{j=0}^i p_j - l, \,\,\,\, \forall i \ge p+1, \nonumber\\
q'_{i+1} = q + K'_i = q + \sum_{j=0}^i p'_j, \,\,\,\, \forall  i \le q+1; \,\,\,\,&\bar{q}'_{i+1} = q + K'_i -l = q + \sum_{j=0}^i p'_j - l, \,\,\,\,\forall i \ge q+1.
\end{eqnarray}
For the sake of convenience, we have defined the new variables $K_i = \sum_{j=0}^i p_i$ and $K'_i = \sum_{j=0}^i p'_i$ which represent the total accumulated momentum exchanges between the hard parton and the nuclear medium.
For such a diagram (Fig. \ref{mntwist}), we may write down the hadronic tensor as,
\begin{eqnarray}
W^{A\mu\nu}_{nmpq} \!\!&=&\!\! \sum_q Q_q^4 e^2 g^{n+m} \frac{1}{N_c} {\rm Tr}\left[\left(\prod_{i=1}^{n} T^{a_i}\right) \left(\prod_{j=m}^{1} T^{a'_j}\right) \right]
\int \frac{d^4l}{(2\pi)^4} (2\pi)\delta(l^2) \int \frac{d^4l_q}{(2\pi)^4} (2\pi)\delta(l_q^2) \nonumber\\\!\!&\times&\!\!
\int d^4y_0 e^{iq\cdot y_0} \left(\prod_{i=1}^{n} \int d^4y_i\right) \left(\prod_{j=1}^{m} \int d^4y'_j \right) \int d^4z \int d^4z'
\nonumber \\\!\!&\times&\!\!
\left(\prod_{i=1}^{p} \int \frac{d^4q_i}{(2\pi)^4} e^{-iq_i\cdot (y_{i-1} - y_i)} \right) \left( \int \frac{d^4q_{p+1}}{(2\pi)^4} e^{-i q_{p+1} \cdot (y_p-z)} e^{-il\cdot z} \int \frac{d^4\bar{q}_{p+1}}{(2\pi)^4} e^{-i\bar{q}_{p+1}\cdot (z-y_{p+1})} \right) \nonumber\\\!\!&\times&\!\!
\left(\prod_{i=p+2}^{n} \int \frac{d^4\bar{q}_i}{(2\pi)^4} e^{-i\bar{q}_i\cdot (y_{i-1} - y_i)} \right) e^{-il_q \cdot (y_{n} - {y'}_{m})}
\left(\prod_{j=q+2}^{m}  \int \frac{d^4\bar{q}'_j}{(2\pi)^4} e^{-i\bar{q}'_j\cdot (y'_j - {y'}_{j-1})}\right)
\nonumber \\ \!\!&\times&\!\!
\left( \int \frac{d^4\bar{q}_{q+1}'}{(2\pi)^4} e^{-i \bar{q}_{q+1}' \cdot (y'_{q+1}-z')} e^{il\cdot z'} \int \frac{d^4q_{q+1}'}{(2\pi)^4} e^{-iq_{q+1}'\cdot (z'-y'_q)} \right)
 \left(\prod_{j=1}^{q}  \int \frac{d^4q'_j}{(2\pi)^4} e^{-iq'_j\cdot (y'_j - {y'}_{j-1})} \right)
\nonumber \\ \!\!&\times&\!\!
\langle A | \bar{\psi}(y_0) \gamma^\mu \left(\prod_{i=1}^{p} \frac{\gamma\cdot q_i}{q_i^2 - i\epsilon} \gamma\cdot A^{a_i}(y_i)\right) \frac{\gamma \cdot q_{p+1}}{q_{p+1}^2 - i \epsilon} \gamma^\alpha \left(\prod_{i=p+1}^{n} \frac{\gamma\cdot \bar{q}_i}{\bar{q}_i^2 - i\epsilon} \gamma\cdot A^{a_i}(y_i)\right) \gamma\cdot l_q  \nonumber\\&\times&
\left(\prod_{j=m}^{q+1} \gamma\cdot A^{a'_j}(y'_j) \frac{\gamma\cdot \bar{q}'_j}{\bar{q'}_j^2 + i \epsilon}\right)
 \gamma^\beta \frac{\gamma \cdot {q'}_{q+1}}{{q'}_{q+1}^2 + i \epsilon}
\left(\prod_{j=q}^{1} \gamma\cdot A^{a_j'}(y'_j) \frac{\gamma\cdot q'_j}{{q'_j}^2 + i \epsilon}\right) \gamma^\nu \psi(0) |A\rangle G_{\alpha\beta}(l) .
\end{eqnarray}
The phase factor associated with the photon insertion locations $z$ and $z'$ may be isolated and it reads $e^{-i(\bar{q}_{p+1} + l - q_{p+1}) \cdot z}  e^{i(\bar{q}'_{p+1} + l - q'_{p+1}) \cdot z'}$.
One may carry out the integration over the locations $z$ and $z'$, and obtain two $\delta$ functions.
These $\delta$ functions can be utilized to perform the integrations over the momenta $\bar{q}_{p+1}$ and $\bar{q}'_{q+1}$, which give us $\bar{q}_{p+1} = q_{p+1} -l$ and $\bar{q}'_{q+1} = q'_{q+1} - l$.
The remainder of this phase factor can be expressed as: $\left(\prod_{i=1}^n e^{-ip_i\cdot y_i} \right) \left(\prod_{j=1}^m e^{ip'_j\cdot y'_j} \right)$.

\begin{figure}[thb]
\includegraphics[width=1.0\textwidth]{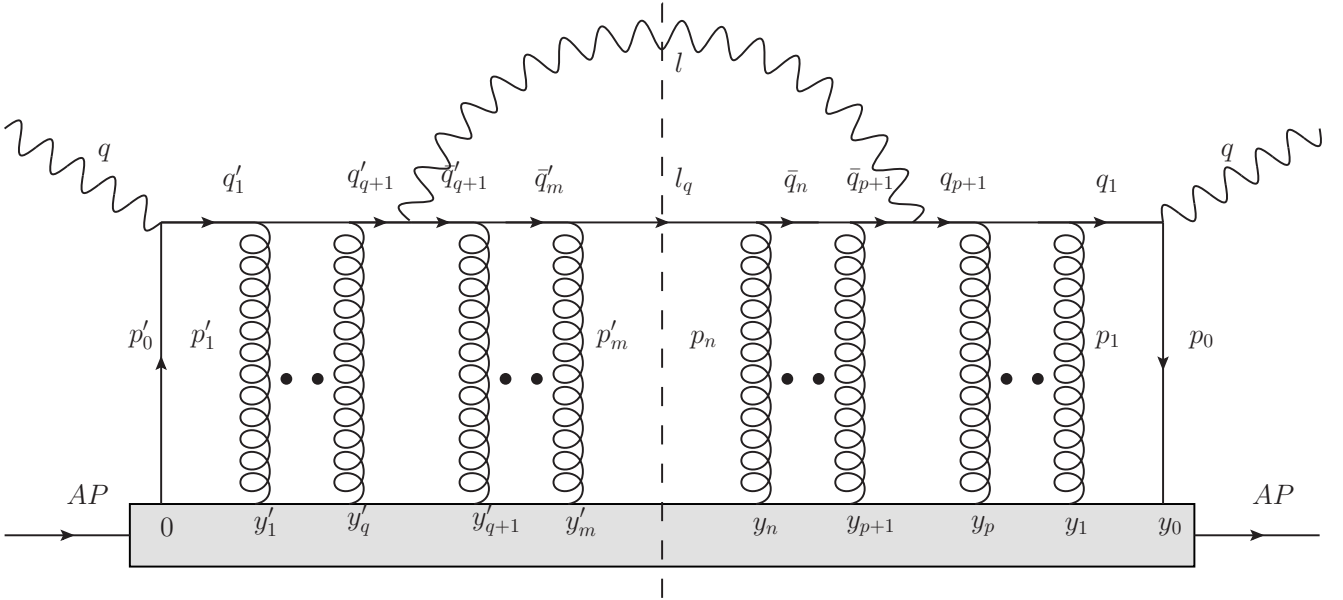}
 \caption{An order of $n+m$ contribution to hadronic tensor $W^{\mu\nu}$ with $n$ gluon insertions in the complex conjugate and $m$ gluon insertions in the amplitude.
} \label{mntwist}
\end{figure}

Similar to the previous section, we may factor out one-nucleon state from the nucleus state and ignore the $(\perp)$-component of the field operators,
\begin{eqnarray}
\langle A | \bar{\psi}(y_0) \gamma^\mu \hat{O} \gamma^\nu \psi(0) | A \rangle
= A C_p^A \langle p | \bar{\psi}(y_0) \frac{\gamma^+}{2} \psi(0) | p \rangle
{\rm Tr} [\frac{\gamma^-}{2} \gamma^\mu \frac{\gamma^+}{2} \gamma^\nu]  {\rm Tr} [\frac{\gamma^-}{2} \langle A| \hat{O} |A\rangle].
\end{eqnarray}
Now we change the integration variables $q_{i+1} \to p_i$ and $\bar{q}'_{j+1} \to p'_j$, and re-introduce the $n$-th momentum $p_n$ by inserting the following momentum conservation identity,
\begin{eqnarray}
\int \frac{d^4p_n}{(2\pi)^4} (2\pi)^4 \delta^4(l+l_q - K_n - q) = 1.
\end{eqnarray}
The $m$-th momentum $p'_m$ in the amplitude is determined by all other momenta as, $p'_m = K_n - K'_{m-1}$.
Since the calculation will be carried out in the light-cone gauge ($A^-=0$) in the Breit frame at very high energy, the dominant component of the vector potential is the forward $(+)$-component. 
In this effort, we approximate $\gamma \cdot A = \gamma^- A^+$, and neglect the contribution from $A_\perp$ component. 
This is due to the fact that the non-zero contribution containing an $A_\perp$ term is of form $(\gamma_\perp \cdot p_{i\perp}) (\gamma_\perp \cdot A_\perp)$, which is two orders smaller than the contribution containing $A^+$ term. 
As a result, the corresponding $\gamma$ matrices have only $(-)$-component.
With these setups, the hadronic tensor now reads,
\begin{eqnarray}
W^{A\mu\nu}_{nmpq} \!\!&=&\!\! \sum_q Q_q^4 e^2 g^{n+m} \frac{1}{N_c} {\rm Tr}\left[\left(\prod_{i=1}^{n} T^{a_i}\right) \left(\prod_{j=m}^{1} T^{a'_j}\right) \right]
\int \frac{d^4l}{(2\pi)^4} (2\pi)\delta(l^2) \int \frac{d^4l_q}{(2\pi)^4} (2\pi)\delta(l_q^2) %\nonumber
\\ \!\!&\times&\!\!
\int d^4y_0 \left(\prod_{i=1}^{n} \int d^4y_i \right) \left(\prod_{j=1}^{m} \int d^4y'_j \right)
\left(\prod_{i=0}^{n-1} \frac{d^4p_i}{(2\pi)^4} \right) \left(\prod_{j=0}^{m-1} \int \frac{d^4p'_j}{(2\pi)^4} \right)
\int \frac{d^4p_n}{(2\pi)^4} (2\pi)^4 \delta^4(l+l_q - K_n - q) \nonumber\\ \!\!&\times&\!\!
\left(\prod_{i=1}^n e^{-ip_i\cdot y_i} \right) \left(\prod_{j=1}^m e^{ip'_j\cdot y'_j} \right)
\left(\prod_{i=1}^{p+1} \frac{1}{q_i^2-i\epsilon} \right) \left(\prod_{i=p+1}^{n} \frac{1}{\bar{q}_i^2-i\epsilon} \right)
\left(\prod_{j=1}^{q+1} \frac{1}{{q'}_j^2+i\epsilon} \right) \left(\prod_{j=q+1}^{m} \frac{1}{\bar{q'}_j^2+i\epsilon} \right) \nonumber \\ \!\!&\times&\!\!
(-g_\perp^{\mu\nu}) A C_p^A \langle p | \bar{\psi}(y_0) \frac{\gamma^+}{2} \psi(0) |p\rangle
\langle A | \left(\prod_{i=1}^{n} A^{+a_i}(y_i)\right) \left(\prod_{j=m}^{1} A^{+a'_j}(y'_j) \right) |A\rangle G_{\alpha\beta}(l)  \nonumber \\ \!\!&\times&\!\!
{\rm Tr}\left[\frac{\gamma^-}{2} \left(\prod_{i=1}^{p} {\gamma\cdot q_i} \gamma^- \right) {\gamma \cdot q_{p+1}} \gamma^\alpha \left(\prod_{i=p+1}^{n} {\gamma\cdot \bar{q}_i} \gamma^-
\right) \gamma\cdot l_q  \left(\prod_{j=m}^{q+1} \gamma^- {\gamma\cdot \bar{q}'_j}\right)
\gamma^\beta {\gamma \cdot {q'}_{q+1}} \left(\prod_{j=q}^{1} \gamma^- {\gamma\cdot q'_j}\right)\right]. \nonumber \ \
\end{eqnarray}
The largest length-enhanced contribution comes from the situation where the maximum number of quark lines are close to on-shell.
For the quark lines before the photon emission,
\begin{eqnarray}
q_{i+1}^2 = (q+K_i)^2 = 2p^+q^-(1+K_i^-/q^-)[-x_B + \bar{x}_i - \bar{x}_{Di}],
\end{eqnarray}
where we have defined the momentum factions,
\begin{eqnarray}
\bar{x}_i = \sum_{j=0}^i x_j = \sum_{j=0}^i \frac{p_j^+}{p^+} = \frac{K_i^+}{p^+}; &&
\bar{x}_{Di} = \sum_{j=0}^i {x}_{Dj} = \frac{K_{i\perp}^2}{2p^+q^-(1+K_i^-/q^-)}.
\end{eqnarray}
For the quark line after photon emission,
\begin{eqnarray}
\bar{q}_{i+1}^2 = (q+K_i-l)^2 = 2p^+q^-(1+K_i^-/q^--y)[-x_B + \bar{x}_i - \bar{x}_{Ci}],
\end{eqnarray}
where we have defined the momentum factions,
\begin{eqnarray}
\bar{x}_{Ci} = \sum_{j=0}^i {x}_{Cj} = x_L(1-y) + \frac{( \mathbf{K}_{i\perp} - \mathbf{l}_\perp)^2}{2p^+q^-(1+K_i^-/q^--y)}.
\end{eqnarray}
Therefore the contribution from all the internal quark line denominators together with the on-shell delta function of the final quark $l_q$ gives,
\begin{eqnarray}
D_q \!\!&=&\!\! \frac{C_q }{(2p^+q^-)^{n+m+3}}
\left( \prod_{i=0}^{p} \frac{1}{-x_B+\bar{x}_i-\bar{x}_{Di}} \right)
\left(\prod_{i=p}^{n-1} \frac{1}{-x_B+\bar{x}_i-\bar{x}_{Ci}} \right)
\left( \prod_{j=0}^{q} \frac{1}{-x_B+{\bar{x}'}_i-{\bar{x}'}_{Di}} \right)
 \nonumber\\
\!\!&\times&\!\! \left(\prod_{j=q}^{m-1} \frac{1}{-x_B+{\bar{x}'}_i-{\bar{x}'}_{Ci}} \right)(2\pi)\delta(-x_B+\bar{x}_n-\bar{x}_{Cn}),
\end{eqnarray}
where the factor $C_q$ stands for,
\begin{eqnarray}
C_q \!\!&=&\!\! \left( \prod_{i=0}^{p} \frac{1}{1+K_i^-/q^-} \right) \left(\prod_{i=p}^{n} \frac{1}{1+K_i^-/q^--y} \right)
\left( \prod_{j=0}^{q} \frac{1}{1+{K'_j}^-/q^-} \right) \left(\prod_{j=q}^{m-1} \frac{1}{1+{K'_j}^-/q^--y} \right) .\ \
\end{eqnarray}
The simplification of the numerators of the quark lines (the trace part) can be done by isolating the largest components of the momentum that are related to various components of the photon polarization sum $G_{\alpha\beta(l)}$. We can decompose $G_{\alpha \beta}$ into the following components,
\begin{eqnarray}
 G^{+-} = G^{-+} = 0; \  G^{++} = \frac{2l^+}{l^-}; \  G^{\perp+} = G^{+\perp} = \frac{l_\perp}{l^-}; \ G_{\perp\perp}^{\alpha\beta} = -g_\perp^{\alpha\beta}.
\end{eqnarray}
The trace combined with the projection $G_{\perp\perp}$ now reads,
\begin{eqnarray}
N_{\perp\perp} \!\!&=&\!\! (-g_\perp^{\alpha\beta}) {\rm Tr} \left[\frac{\gamma^-}{2}\right. \left(\prod_{i=1}^p \gamma^+(q^-+K_{i-1}^-)\gamma^-\right)
\left\{\gamma^+(q^-+K_p^-) + \mathbf{\gamma}_\perp \cdot \mathbf{K}_{p \perp}\right\}  \nonumber \\
\!\!&\times&\!\! \gamma_\perp^\alpha \left\{\gamma^+(q^-+K_p^--l^-) + \mathbf{\gamma}_\perp \cdot (\mathbf{K}_{p \perp}-\mathbf{l}_\perp)\right\}
\gamma^-\left(\prod_{i=p+2}^n \gamma^+(q^-+K_{i-1}^--l^-)\gamma^-\right)\nonumber \\
\!\!&\times&\!\! (\gamma^+l_q^-) \left(\prod_{j=m}^{q+2} \gamma^-\gamma^+(q^-+{K'}_{j-1}^--l^-)\right)\gamma^-
\left\{\gamma^+(q^-+{K'_q}^--l^-) + \mathbf{\gamma}_\perp \cdot ({\mathbf{K}'}_{q\perp}-\mathbf{l}_\perp)\right\} \nonumber \\
\!\!&\times&\!\! \gamma_\perp^\beta\left\{\gamma^+(q^--{K'_q}^-) + \mathbf{\gamma}_\perp \cdot
{\mathbf{K}'}_{q\perp}\right\} \left(\prod_{j=q}^{1} \gamma^-\gamma^+(q^-+{K'}_{j-1}^-)\left. \right)\right].
\end{eqnarray}
Here the transverse components of the quark momentum in the numerators of the propagators are only retained for those terms immediately before and after
the photon insertion; these terms yield leading contributions.
Using the anti-commutation relations of the $\gamma$ matrices, the above expression may be simplified as:
\begin{eqnarray}
N_{\perp\perp} = \frac{2(2q^-)^{n+m+1}}{C_q}
 \left[\frac{\mathbf{K}_{p \perp} \cdot {\mathbf{K}'}_{q\perp}}{(1+K_p^-/q^-)(1+{K'_q}^-/q^-)} + \frac{(\mathbf{K}_{p \perp} -\mathbf{l}_\perp) \cdot ({\mathbf{K}'}_{q\perp}-\mathbf{l}_\perp)}{(1+K_p^-/q^--y)(1+{K'_q}^-/q^--y)}\right].
\end{eqnarray}
Using a similar procedure as above, the terms in the trace originating from the projection $G^{++}$ gives,
\begin{eqnarray}
N_{++} = \frac{4(2q^-)^{n+m+1}}{C_q} \frac{l_\perp^2}{y^2}.
\end{eqnarray}
The contribution from the terms associated with the projections $G^{\perp +}$ and $G^{+\perp}$ reads,
\begin{eqnarray}
N_{\perp+} + N_{+\perp} = -\frac{2(2q^-)^{n+m+1}}{C_q} \frac{1}{y}
\left[\frac{\mathbf{K}_{p \perp} \cdot \mathbf{l}_{\perp}}{(1+K_p^-/q^-)} + \frac{{\mathbf{K}'}_{q \perp} \cdot \mathbf{l}_{\perp}}{(1+{K'_q}^-/q^-)} + \frac{(\mathbf{K}_{p \perp} -\mathbf{l}_\perp) \cdot \mathbf{l}_{\perp}}{(1+K_p^-/q^--y)} + \frac{({\mathbf{K}'}_{q \perp} -\mathbf{l}_\perp) \cdot \mathbf{l}_{\perp}}{ (1+{K'_q}^-/q^--y)}\right] . \ \
\end{eqnarray}
Combining various parts with all of the above simplifications, the hadronic tensor now takes the following form,
\begin{eqnarray}
W^{A\mu\nu}_{nmpq} \!\!&=&\!\! \sum_q Q_q^4 e^2 g^{n+m} \frac{1}{N_c} {\rm Tr}\left[\left(\prod_{i=1}^{n} T^{a_i}\right) \left(\prod_{j=m}^{1} T^{a'_j}\right) \right]
\int \frac{d^4l}{(2\pi)^4} (2\pi)\delta(l^2) \int \frac{d^4l_q}{(2\pi)^4}   (2\pi)^4 \delta^4(l+l_q - K_n - q) \nonumber\\\!\!&\times&\!\!
\int d^4y_0 \left(\prod_{i=1}^{n} \int d^4y_i \right) \left(\prod_{j=1}^{m} \int d^4y'_j \right)
\left(\prod_{i=0}^{n-1} \frac{d^3p_i dx_i}{(2\pi)^4} \right) \left(\prod_{j=0}^{m-1} \int \frac{d^3p'_jdx'_j}{(2\pi)^4} \right)
\left( \int \frac{d^3p_n dx_n}{(2\pi)^4} \right) \nonumber\\ \!\!&\times&\!\!
\left(\prod_{i=0}^n e^{-i\mathbf{p}_i\cdot \mathbf{y}_i} e^{-ix_ip^+y_i^-} \right) \left(\prod_{j=0}^m e^{i{\mathbf{p}'}_j\cdot {\mathbf{y}'}_j e^{ix'_jp^+{y'_i}^-} } \right)
\left( \prod_{i=0}^{p} \frac{1}{-x_B+\bar{x}_i-\bar{x}_{Di}-i\epsilon} \right) \left(\prod_{i=p}^{n-1} \frac{1}{-x_B+\bar{x}_i-\bar{x}_{Ci}-i\epsilon} \right) \nonumber\\\!\!&\times&\!\!
\left( \prod_{j=0}^{q} \frac{1}{-x_B+{\bar{x}'}_i-{\bar{x}'}_{Di}+i\epsilon} \right) \left(\prod_{j=q}^{m-1} \frac{1}{-x_B+{\bar{x}'}_i-{\bar{x}'}_{Ci}+i\epsilon} \right) (2\pi)\delta(-x_B+\bar{x}_n-\bar{x}_{Cn})\nonumber\\
\!\!&\times&\!\!  (-g_\perp^{\mu\nu}) A C_p^A \langle p | \bar{\psi}(y_0) \frac{\gamma^+}{2} \psi(0) |p\rangle
 \langle A | \left(\prod_{i=1}^{n} A^{+a_i}(y_i)\right) \left(\prod_{j=m}^{1} A^{+a'_j}(y'_j) \right) |A\rangle  \nonumber \\ &&
\nonumber\\ \!\!&\times&\!\!
\frac{2}{(2p^+q^-)^2}
\frac{1 + \left(1 - \frac{y}{1+{K_p^-}/{q^-}} \right) \left(1 -  \frac{y }{1+{{K'_q}^-}/{q^-}} \right)}{y^2\left(1 - \frac{y}{1+{K_p^-}/{q^-}} \right) \left(1 -  \frac{y }{1+{{K'_q}^-}/{q^-}} \right)}
\left(\mathbf{l}_\perp - \frac{y }{1+\frac{K_p^-}{q^-}} \mathbf{K}_{p\perp} \right)\cdot \left(\mathbf{l}_\perp - \frac{y }{1+\frac{{K'_q}^-}{q^-}} {\mathbf{K}'}_{q\perp} \right). \ \
\end{eqnarray}
%
%\end{widetext}
%
Here we have changed the integration variables $p_i^+ \to x_i = p_i^+/p^+$ and ${p'_j}^+ \to x'_j = {p'_j}^+/p^+$.
This renders a factor of $(p^+)^{n+m+1}$ to cancel that in the expression of $D_q$.
Also for convenience, we have used the three-vector notations for momentum and space variables: $\mathbf{p} = (p^-, \mathbf{p}_\perp)$ and $\mathbf{y} = (y^+, \mathbf{y}_\perp)$, and their dot product $\mathbf{p}\cdot \mathbf{y} = p^-y^+ - \mathbf{p}_\perp \cdot \mathbf{y}_\perp$.

Now we do the integrations over the momentum fractions $x_i$ and $x'_j$. The phase factors related to them read as,
\begin{eqnarray}
\Gamma^+ = \prod_{i=0}^n e^{-ix_ip^+(y_i^--{y'_m}^-)} \prod_{j=0}^{m-1} e^{ix'_jp^+({y'_j}^--{y'_m}^-)}\!\!,
\end{eqnarray}
where we have used the overall momentum conservation $p_m = \sum_{i=0}^n p_i - \sum_{j=0}^{m-1} p'_j$.
Further using the on-shell delta function for the last quark line $(2\pi) \delta(-x_B+\bar{x}_n - \bar{x}_{Cn})$, we can perform the integration over the last momentum fraction $x_n$. Then the phase factor can be rearranged as
\begin{eqnarray}
\Gamma^+ = e^{-i(x_B+\bar{x}_{Cn})p^+y_n^-} \prod_{i=0}^{n-1} e^{-ix_ip^+(y_i^--y_n^-)}
e^{i(x_B+\bar{x}'_{Cm})p^+{y'_m}^-} \prod_{j=0}^{m-1} e^{ix'_jp^+({y'_j}^--{y'_m}^-)}  = \Gamma_n^+ \Gamma_m^+ .
\end{eqnarray}
Here we use $\Gamma_n^+$ and $\Gamma_m^+$ to denote the factors associated with $x_i$ ($1\le i\le n$) and $x'_j$ ($1\le j\le m$) integrals, respectively.
The remaining integration over $x_i$ and $x'_j$ may be performed using the technique of contour integration.

We start from the propagators adjacent to the cut and proceed to the propagators adjacent to the photon vertices. In the following, we write down the details for the $x_i$ integration; the $x'_j$ integration is completely analogous. The first integration is over the momentum fraction $x_{n-1}$.
Isolating the piece related to $x_{n-1}$ integral, we can perform the integral by closing the contour of $x_{n-1}$ with a counter-clockwise semi-circle in the upper half of the complex plane,
\begin{eqnarray}
\int \frac{dx_{n-1}}{2\pi} \frac{e^{-ix_{n-1}p^+(y_{n-1}^--y_n^-)}}{-x_B + \bar{x}_{n-1} - \bar{x}_{Cn-1} - i \epsilon}  = i \theta(y_n^- - y_{n-1}^-) e^{-i(x_B + \bar{x}_{Cn-1} - \bar{x}_{n-2})p^+(y_{n-1}^- - y_n^-)}. \ \
\end{eqnarray}
Here the $\theta$-function represents the physical effect that the quark line is propagating from $y_{n-1}^-$ to $y_n^-$.
Correspondingly, the phase factor combined with these from the contour integration becomes,
\begin{eqnarray}
\Gamma_n^+ \to i\theta(y_n^- - y_{n-1}^-) e^{-i{x}_{Cn}p^+ y_n^-}  e^{-i(x_B+\bar{x}_{Cn-1})p^+y_{n-1}^-}
\left(\prod_{i=0}^{n-2} e^{-ix_ip^+(y_i^--y_n^-)} \right).
\end{eqnarray}
The remaining longitudinal momentum fraction $x_i$ integrals can be done in a similar way until we reach the integration involving $x_{p+1}$. The phase factor up to this step reads,
\begin{eqnarray}
\Gamma_n^+ \to \left(\prod_{i=p+2}^n i\theta(y_i^- - y_{i-1}^-) e^{-i{x}_{Ci} p^+ y_i^-} \right) e^{-i(x_B+\bar{x}_{Cp+1})p^+y_{p+1}^-}
 \left(\prod_{i=0}^{p} e^{-ix_ip^+(y_i^--y_{p+1}^-)} \right).
\end{eqnarray}
Now we perform the integration on $x_p$. Note that around the photon insertion, there are two propagators, either of which can be evaluated to obtain the on-shell conditions for $x_p$. Performing the contour integration for $x_p$, we obtain
\begin{eqnarray}
&& \int \frac{dx_p^+}{2\pi}  \frac{e^{-ix_p p^+(y_p^--y_{p+1}^-)}}{(-x_B + \bar{x}_p - \bar{x}_{Cp} - i \epsilon)(-x_B + \bar{x}_p - \bar{x}_{Dp} - i \epsilon)} \nonumber\\
&&= {i\theta(y_{p+1}^--y_p^-)   e^{-ix_B p^+(y_p^--y_{p+1}^-)}}
\frac{ e^{-i\bar{x}_{Cp}(y_p^- - y_{p+1}^-)} - e^{-i\bar{x}_{Dp}(y_p^- - y_{p+1}^-)}} {\bar{x}_{Cp} - \bar{x}_{Dp}} e^{i\bar{x}_{p-1}^+p^+(y_p^--y_{p+1}^-)}. \ \
\end{eqnarray}
The above two contributions represent the fact that the leading length enhancement arises when at least one of the propagators is close to on-shell.
The phase factor up to this step becomes,
\begin{eqnarray}
\Gamma_n^+ \to \left(\prod_{i=p+1}^n i\theta(y_i^- - y_{i-1}^-) e^{-i{x}_{Ci} p^+ y_i^-} \right) e^{-i(x_B+\bar{x}_{Dp})p^+y_p^-}
\left(\prod_{i=0}^{p-1} e^{-ix_ip^+(y_i^--y_p^-)} \right) \frac{e^{-i\delta\bar{x}_{Dp} p^+ y_p^-} - e^{-i\delta\bar{x}_{Dp} p^+ y_{p+1}^-}} {\delta\bar{x}_{Dp}}, \ \
\end{eqnarray}
where we have defined the variables $\delta\bar{x}_{Di} = \bar{x}_{Ci} - \bar{x}_{Di}$ for convenience (the definition of $\bar{x}_{Di}$ is extended to the regime $i > p+1$).
The remaining integrations over the propagators after the photon insertion are similar.
After performing the integration over all momentum fractions $x_i$ in the complex conjugate, the phase factor becomes
\begin{eqnarray}
\Gamma_n^+ \to \left(\prod_{i=p+1}^n i\theta(y_i^- - y_{i-1}^-) e^{-i{x}_{Ci} p^+ y_i^-} \right)
\left(\prod_{i=1}^p i\theta(y_i^- - y_{i-1}^-) e^{-i{x}_{Di} p^+ y_i^-} \right)
e^{-ix_Bp^+y_0^-} \frac{e^{-i\delta\bar{x}_{Dp} p^+ y_p^-} - e^{-i\delta\bar{x}_{Dp} p^+ y_{p+1}^-}} {\delta\bar{x}_{Dp}}. \ \
\end{eqnarray}

The integration over the momentum faction $x'_j$ in the amplitude (to get $\Gamma_m^+$) is completely analogous, except an factor of $(-i)$ instead of $i$ associated with the $\theta$ function, which originates from the contour integral which we perform by closing the contour of $x_{n-1}$ with a clockwise semi-circle in the upper half of the complex plane.
After carrying out the integration over the quark lines, we obtain the hadronic tensor as,
\begin{eqnarray}
W^{A\mu\nu}_{nmpq}  \!\!&=&\!\! \sum_q Q_q^4 e^2 g^{n+m} \frac{1}{N_c} {\rm Tr}\left[\left(\prod_{i=1}^{n} T^{a_i}\right) \left(\prod_{j=m}^{1} T^{a'_j}\right) \right]
\int \frac{d^4l}{(2\pi)^4} (2\pi)\delta(l^2) \int \frac{d^4l_q}{(2\pi)^4}   (2\pi)^4 \delta^4(l+l_q - K_n - q) %\nonumber
\\\!\!&\times&\!\!
\int d^4y_0 \left(\prod_{i=1}^{n} \int d^4y_i \right) \left(\prod_{j=1}^{m} \int d^4y'_j \right)
\left(\prod_{i=0}^{n-1} \frac{d^3p_i}{(2\pi)^3} \right) \left(\prod_{j=0}^{m-1} \int \frac{d^3p'_j}{(2\pi)^3} \right)
\int \frac{d^3p_n}{(2\pi)^3}  \left(\prod_{i=0}^n e^{-i\mathbf{p}_i\cdot \mathbf{y}_i} \right) \left(\prod_{j=0}^m e^{i{\mathbf{p}'}_j\cdot {\mathbf{y}'}_j } \right) \nonumber\\\!\!&\times&\!\!
e^{-ix_Bp^+y_0^-}  \left(\prod_{i=1}^n i\theta(y_i^- - y_{i-1}^-) e^{-i{x}_{Di} p^+ y_i^-} \right)
\left(\prod_{j=1}^m (-i)\theta({y'_j}^- - {y'}_{j-1}^-) e^{i{x_D}'_j p^+ {y'}_j^-}\right) \nonumber\\ \!\!&\times&\!\!
(-g_\perp^{\mu\nu}) A C_p^A \langle p | \bar{\psi}(y_0) \frac{\gamma^+}{2} \psi(0) |p\rangle
 \langle A | \left(\prod_{i=1}^{n} A^{+a_i}(y_i)\right) \left(\prod_{j=m}^{1} A^{+a'_j}(y'_j) \right) |A\rangle  \nonumber \\ \!\!&\times&\!\!
 \left(\prod_{i=p+1}^n e^{-i\delta{x}_{Di} p^+ y_i^-} \right)
 \left(\prod_{j=q+1}^n e^{i\delta{x_D}'_j p^+ {y'_j}^-} \right)
 %\nonumber \\ \!\!&\times&\!\!
 \left(e^{-i\delta\bar{x}_{Dp} p^+ y_p^-} - e^{-i\delta\bar{x}_{Dp} p^+ y_{p+1}^-}\right)
\left(e^{i\delta\bar{x}'_{Dq} p^+ {y'_p}^-} - e^{i\delta\bar{x}'_{Dq} p^+ {y'}_{q+1}^-}\right)
\nonumber\\ \!\!&\times&\!\!
\frac{1}{\delta\bar{x}_{Dp} \delta\bar{x}'_{Dq}}
\frac{2}{(2p^+q^-)^2}
\frac{1 + \left(1 - \frac{y}{1+{K_p^-}/{q^-}} \right) \left(1 -  \frac{y }{1+{{K'_q}^-}/{q^-}} \right)}{y^2\left(1 - \frac{y}{1+{K_p^-}/{q^-}} \right) \left(1 -  \frac{y }{1+{{K'_q}^-}/{q^-}} \right)}
\left(\mathbf{l}_\perp - \frac{y }{1+\frac{K_p^-}{q^-}} \mathbf{K}_{p\perp} \right)\cdot \left(\mathbf{l}_\perp - \frac{y }{1+\frac{{K'_q}^-}{q^-}} {\mathbf{K}'}_{q\perp} \right), \nonumber%\\
\end{eqnarray}
where we have defined the variable $\delta{x}_{Di} = {x}_{Ci} - {x}_{Di}$ for convenience (the definition of ${x}_{Di}$ is extended to the regime $i > p+1$).
\end{widetext}

\section{Sum over photon production points}
~\label{sum-sect}

In the previous section, we have derived the single photon radiation spectrum from a hard quark jet which experiences multiple scattering off the gluon field within the nucleus. The expression is quite general in that we have not made any assumption regarding the nature of the nuclear states.
In this section, we will carry out the resummation over different photon insertion locations.
Here we only consider the case with $n=m$, i.e.,the number of scatterings in both the amplitude and the complex conjugate are the same.
The case with $n\ne m$ either constitutes higher twist nucleon matrix elements or represents the unitarity corrections to the terms where the quark experiences scatterings $\min(n,m)$ times \cite{Majumder:2007hx, Majumder:2007ne}.
The case with $p=q$ represents the squares of the amplitude, and the case with $p\ne q$ represents the LPM interference terms.

For $2n$ ($n=m$) gluon insertions, we first simplify the matrix elements of the gluon vector potentials in the nuclear state.
In this work, we assume that the nucleus may be approximated by a weakly interacting homogenous gas of nucleons.
Such approximation is reasonable in very high energy limit, since nucleons are traveling in straight lines and are almost independent of each other over the time interval of the interaction of the hard probe with the medium due to time dilation.
As a consequence, the expectation of field operators in the nucleus states may be decomposed into a product of the expectations in the nucleon states.
Due to the fact that a nucleon is a color singlet, any combination of quark or gluon field insertions should be restricted to a color singlet.
Therefore, the first non-zero and the largest contribution comes from the terms where $2n$ gluons are divided into singlet pairs in separate nucleon states.

With the help of the time-ordered products of $\theta$ functions, the expectation of the $2n$ gluon operators in the nucleus state may be decomposed as,
\begin{widetext}
\begin{eqnarray}
 \langle A | \left(\prod_{i=1}^{n} A^{+a_i}(y_i)\right) \left(\prod_{j=n}^{1} A^{+a'_j}(y'_j) \right) |A\rangle
= \left(\frac{\rho}{2p^+}\right)^n \left(\prod_{i=1}^{n} \frac{\delta_{a_i a'_i}}{N_c^2 - 1} \langle p| A^+(y_i) A^+(y'_i) | p\rangle\right),
\end{eqnarray}
where we have averaged over the colors of gluon field operators. Note that the quark operators has been factorized out in the previous section.
With such decomposition, we can easily evaluate the trace for the color matrices,
\begin{eqnarray}
\frac{1}{N_c(N_c^2 - 1)^n} {\rm Tr}\left[\left(\prod_{i=1}^{n} T^{a_i}\!\right) \!\! \left(\prod_{i=n}^1 T^{a'_i}\!\right) \right] = \left( \frac{C_F}{N_c^2 - 1}\right)^n.
\end{eqnarray}
We may further change the integral variables $\mathbf{y}$ and $\mathbf{y}'$,
\begin{eqnarray}
\mathbf{Y}_i = (\mathbf{y}_i + \mathbf{y}'_i) /2;  &&  \delta\mathbf{y}_i = \mathbf{y}_i - \mathbf{y}'_i.
\end{eqnarray}
Using the translational invariance of the correlation functions,
\begin{eqnarray}
\langle A(y_i^-, \mathbf{y}_i) A({y'_i}^-, \mathbf{y}'_i)  \rangle = \langle  A(y_i^-, \delta \mathbf{y}_i) A({y'_i}^-, \mathbf{0})  \rangle,
\end{eqnarray}
we may carry out the integration for the phase factor,
\begin{eqnarray}
\int d^3y_i \int d^3y'_i e^{-i\mathbf{p}_i \cdot \mathbf{y}_i} e^{i\mathbf{p}'_i\cdot \mathbf{y}'_i}
= (2\pi)^3 \delta^3(\mathbf{p}_i - \mathbf{p}'_i) \int d^3\delta y_i e^{-i\mathbf{p}_i\cdot \delta\mathbf{y}_i}.
\end{eqnarray}
The $\delta$ function means that each pair of gluon field insertions in each nucleon states carry the same momentum, so we may use the $\delta$ function to carry out the integration over the momentum $\mathbf{p}'_i$.
Putting various parts together, the hadronic tensor now reads,
\begin{eqnarray}
W^{A\mu\nu}_{nnpq} \!\!&=&\!\! (-g_\perp^{\mu\nu}) A C_p^A \sum_q Q_q^4 e^2 g^{2n} \left( \frac{C_F}{N_c^2 - 1}\right)^n \left(\frac{\rho}{2p^+}\right)^n
\int \frac{d^4l}{(2\pi)^4} (2\pi)\delta(l^2) \int \frac{d^4l_q}{(2\pi)^4}   (2\pi)^4 \delta^4(l+l_q - K_n - q) %\nonumber
\\\!\!&\times&\!\!
\int dy_0^- d^3y_0 \int \frac{d^3p_0}{(2\pi)^3} e^{-i\mathbf{p}_0 \cdot \mathbf{y}_0} e^{-ix_B p^+y_0^-}  \langle p | \bar{\psi}(y_0) \frac{\gamma^+}{2} \psi(0) |p\rangle \left(\prod_{i=1}^{n} \int dy_i^- d{y'_i}^- \theta(y_i^- - y_{i-1}^-) \theta({y'_i}^- - {y'}_{i-1}^-)  \right) \nonumber\\\!\!&\times&\!\!
\left(\prod_{i=1}^n e^{-i{x}_{Di} p^+ y_i^-} e^{i{x'}_{Di} p^+ {y'}_i^-}\right)
\left(\prod_{i=1}^{n} \int d^3\delta y_i \frac{d^3p_i}{(2\pi)^3} e^{-i\mathbf{p}_i\cdot \delta \mathbf{y}_i} \right)
 \left(\prod_{i=1}^{n} \langle p | A^+(y_i^-, \delta \mathbf{y}_i) A^+({y'_i}^-, \mathbf{0})  |p\rangle \right)
\nonumber \\ \!\!&\times&\!\!
 \left(\prod_{i=p+1}^n e^{-i\delta{x}_{Di} p^+ y_i^-} \right)
 \left(\prod_{i=q+1}^n e^{i\delta{x'}_{Di} p^+ {y'_i}^-} \right)
 %\nonumber\\ \!\!&\times&\!\!
 \left({e^{-i\delta\bar{x}_{Dp} p^+ y_p^-} - e^{-i\delta\bar{x}_{Dp} p^+ y_{p+1}^-}}\right)
\left({e^{i\delta\bar{x}'_{Dq} p^+ {y'_p}^-} - e^{i\delta\bar{x}'_{Dq} p^+ {y'}_{q+1}^-}}\right)
\nonumber\\ \!\!&\times&\!\!
\frac{1}{\delta\bar{x}_{Dp} \delta\bar{x}'_{Dq}}
\frac{2}{(2p^+q^-)^2}
\frac{1 + \left(1 - \frac{y}{1+{K_p^-}/{q^-}} \right) \left(1 -  \frac{y }{1+{{K'_q}^-}/{q^-}} \right)}{y^2\left(1 - \frac{y}{1+{K_p^-}/{q^-}} \right) \left(1 -  \frac{y }{1+{{K'_q}^-}/{q^-}} \right)}
\left(\mathbf{l}_\perp - \frac{y }{1+\frac{K_p^-}{q^-}} \mathbf{K}_{p\perp} \right)\cdot \left(\mathbf{l}_\perp - \frac{y }{1+\frac{{K'_q}^-}{q^-}} {\mathbf{K}'}_{q\perp} \right). \nonumber %\\
\end{eqnarray}

Up to this point, we have retained all orders in terms of the form $K_{\perp}/l_{\perp}$, i.e., there can be any relation between the accumulated transverse momentum of the hard quark and that of the radiated photon. In the remainder of this paper, we will consider the case of hard photon radiation where $p_{\perp} / l_{\perp} \sim \lambda \ll 1$. As we will demonstrate below, in this limit, it will become possible to derive closed form expressions for the sum over photon emission points. 
To perform the summation of photon insertion points $p$ and $q$, we recall the expression of $\bar{x}_{Dp}$ and $\bar{x}_{Cp}$ which appear in both the phases and matrix elements,
\begin{eqnarray}
\bar{x}_{Dp} = \frac{K_{p\perp}^2}{2p^+q^-(1+K_p^-/q^-)}; && \bar{x}_{Cp} = \frac{l_\perp^2}{2p^+q^-y} + \frac{(\mathbf{K}_{p\perp}-\mathbf{l}_\perp)^2}{2p^+q^-(1+K_p^-/q^- - y)}.
\end{eqnarray}
The expressions for $\bar{x_D}'_q$ and $\bar{x_C}'_q$ are completely analogous. The difference between these two variables gives,
\begin{eqnarray}
\delta\bar{x}_{Dp} = \bar{x}_{Cp} - \bar{x}_{Dp}
= \frac{\left(\mathbf{l}_\perp - \frac{y }{1+{K_p^-}/{q^-}} \mathbf{K}_{p\perp} \right)^2}{2p^+q^-y\left(1-\frac{y}{1+K_p^-/q^-}\right)}
\approx x_L \left(1-\frac{y}{1-y} \frac{K_p^-}{q^-} - 2y \frac{\mathbf{l}_\perp \cdot \mathbf{K}_{p\perp}}{l_\perp^2} \right).
\end{eqnarray}
Here we only keep terms up to linear in $K_p^-/q^-$ and $K_{p_\perp}/l_\perp$. Using the same procedure, the last line of the hadronic tensor (which we denote as $\delta T_{pq}$) may be simplified as,
\begin{eqnarray}
\label{linear_approximation}
\delta T_{pq} = \frac{2yP(y)}{l_\perp^2}
\left[1 + \frac{y(1-y)}{1+(1-y)^2} \frac{K_p^-+{K'_q}^-}{q^-} + y \frac{\mathbf{l}_\perp \cdot (\mathbf{K}_{p\perp}+\mathbf{K}'_{q\perp})}{l_\perp^2} \right].
\end{eqnarray}
To simplify the phase factor, we note that
\begin{eqnarray}
{x}_{Di} = \bar{x}_{Di} - \bar{x_D}_{i-1} \approx 0; & {x}_{Ci} = \bar{x}_{Ci} - \bar{x_C}_{i-1} \approx x_L \left(-y \frac{K_p^-}{q^-} - 2y \frac{\mathbf{l}_\perp \cdot \mathbf{K}_{p\perp}}{l_\perp^2}\right); & \delta {x}_{Di} = {x}_{Ci} - {x}_{Di} \approx {x}_{Ci}.
\end{eqnarray}
One may see that $\delta {x}_{Di} \ll \delta\bar{x}_{Di}$. If we only keep the leading term and ignore the terms in higher order $K_p^-/q^-$, ${K'_q}^-/q^-$ and $K_{p\perp}/l_\perp$, $K'_{q\perp}/l_\perp$, the phase factor can be simplified as
\begin{eqnarray}
\left({e^{-ix_L p^+ y_p^-} - e^{-ix_L p^+ y_{p+1}^-}}\right)
\left({e^{ix_L p^+ {y'_q}^-} - e^{ix_L p^+ {y'}_{q+1}^-}}\right) = g_p g_q.
\end{eqnarray}
Here we use $g_p$ and $g_q$ to represent the phase factors associated with the complex conjugate and the amplitude, respectively.
With these simplifications, we may perform the sum over the photon insertion locations $p$ and $q$,
\begin{eqnarray}
\sum_{p=0}^{n} \sum_{q=0}^n g_p g_q \delta T_{pq} = \frac{2yP(y)}{l_\perp^2} e^{-ix_Lp^+y_0^-} \left[1 + \frac{y(1-y)}{1+(1-y)^2} \sum_{k=1}^n \frac{p_k^-}{q^-} F_k + y \sum_{k=1}^n \frac{\mathbf{l}_\perp \cdot \mathbf{p}_{k\perp}}{l_\perp^2} F_k \right],
\end{eqnarray}
where $F_k$ stands for
\begin{eqnarray}
F_k = {e^{-ix_L p^+ (y_k^--y_0^-)} + e^{ix_L k^+ {y'_p}^-}}  = {e^{-ix_L p^+ (Y_k^-+\delta y_k^-/2 + y_0^-)} + e^{ix_L p^+ (Y_k^- - \delta y_k^-/2)}}.
\end{eqnarray}
In the last expression, we have made the variable change from $y_i$ and $y'_i$ to $Y_i^-$ and $\delta y_i^-$,
\begin{eqnarray}
Y_i^- = (y_i^- + {y'_i}^-)/2, \delta y_i^- = y_i^- - {y'_i}^-.
\end{eqnarray}
Using the the fact that $Y_k^-$ is the location of the photon insertion points which can span over the nucleus size, while $y_0^-$ and $\delta y_k^-$ are confined within the nucleon size, we may drop $y_0^-$ and $\delta y_k^-$ as compared to $Y_k^-$ in the expression of $F_k$ and obtain the following:
\begin{eqnarray}
F_k \approx 2 \cos(x_L p^+ Y_k^-).
\end{eqnarray}
The hadronic tensor now reads,
\begin{eqnarray}
W^{A\mu\nu}_{nn} \!\!&=&\!\! (-g_\perp^{\mu\nu}) A C_p^A \sum_q Q_q^4 \frac{\alpha_e}{2\pi}
\int dy P(y) \int \frac{d^2l_\perp}{\pi l_\perp^2} \int{d^3l_q}
\int dy_0^- e^{-i(x_B+x_L)p^+y_0^-}  \langle p | \bar{\psi}(y_0^-) \frac{\gamma^+}{2} \psi(0) |p\rangle \nonumber\\\!\!&\times&\!\!
\frac{1}{n!} \left(\prod_{i=1}^{n} \int dY_i^- \int d\delta y_i^- \left(g^2\frac{C_F}{N_c^2 - 1}\frac{\rho}{2p^+}\right) \int d^3\delta y_i \frac{d^3p_i}{(2\pi)^3} e^{-i\mathbf{p}_i\cdot \delta \mathbf{y}_i} \langle p | A^+(\delta y_i^-, \delta \mathbf{y}_i) A^+(0)  |p\rangle \right)
\nonumber \\ \!\!&\times&\!\!
\left[1 + \frac{y(1-y)}{1+(1-y)^2} \sum_{p=1}^n \frac{p_p^-}{q^-} F_k + y \sum_{p=1}^n \frac{\mathbf{l}_\perp \cdot \mathbf{p}_{p\perp}}{l_\perp^2} F_k \right] \delta^3(l+l_q - K_n - q).
\end{eqnarray}
\end{widetext}

\section{Resummation of multiple scatterings}
~\label{resum-sect}

In the previous section, we have performed the resummation over different photon production points for a given number of scatterings experienced by the hard parton.
In this section, we will perform the resummation over the number of multiple scatterings.
With the assumption of small momentum exchange in each of the multiple scatterings, we may expand the hard part $H(\mathbf{p}_i)$ of the matrix elements, the last line in the hadronic tensor, as a series of the Taylor expansion in the three-dimensional momenta $\mathbf{p}_i = (p_i^-, \mathbf{p}_{i\perp})$ around $\mathbf{p}_i \to 0$. Keeping the expansion up to the second order derivative in both longitudinal and transverse momenta, we obtain:
\begin{widetext}
\begin{eqnarray}
H = \left(\prod_{i=1}^{n} \left[1 + p_i^\alpha \frac{\partial}{\partial p_i^\alpha} + \frac{1}{2} p_i^\alpha p_i^\beta \frac{\partial}{\partial p_i^\alpha} \frac{\partial}{\partial p_i^\beta} + \cdots \right] \right) H|_{\mathbf{p}_1 \cdots \mathbf{p}_n = 0}. \ \
\end{eqnarray}
Here the values of $\alpha$ and $\beta$ take ``$-$'' or ``$\perp$".
In the above expansion, the zeroth order terms (with no momentum derivative) represent the case where the exchanged gluons carry zero momenta.
These terms correspond to the gauge corrections to the diagrams with lower number of scatterings which carry nonzero momenta.
They should be included in the gauge invariant definition of the jet transport coefficients and will not be considered further.
Terms with momentum gradients higher than the second order are neglected in the above equation and should be straightforward to include.
They correspond to higher moments of the momentum distribution of the exchanged gluons.

Performing the integration by parts over $p_i$, we may use the following substitutions,
\begin{eqnarray}
&&\langle p| A^+(\delta \mathbf{y}_i) A^+(0)|p \rangle e^{-\mathbf{p}_i \cdot \delta\mathbf{y}_i} p_i^\alpha \frac{\partial}{\partial p_i^\alpha} =   e^{-\mathbf{p}_i \cdot \delta\mathbf{y}_i} (-i) \langle p| \partial^\alpha A^+(\delta \mathbf{y}_i) A^+(0)|p \rangle \frac{\partial}{\partial p_i^\alpha},  \\
&&\langle p| A^+(\delta \mathbf{y}_i) A^+(0)|p \rangle e^{-\mathbf{p}_i \cdot \delta\mathbf{y}_i} p_i^\alpha p_i^\beta \frac{\partial}{\partial p_i^\alpha} \frac{\partial}{\partial p_i^\beta} = e^{-\mathbf{p}_i \cdot \delta\mathbf{y}_i} \langle p| \partial^\alpha A^+(\delta \mathbf{y}_i) \partial^\beta A^+(0)|p \rangle \frac{\partial}{\partial p_i^\alpha} \frac{\partial}{\partial p_i^\beta}.
\end{eqnarray}
Now since the Taylor expansion imposes the condition $\mathbf{p}_i \to 0$ on the hard part $H$, it no longer has functional dependence on $\mathbf{p}_i$.
Therefore, we may perform the integrations for $\mathbf{p}_i$ and $\delta\mathbf{y}_i$. With the above steps, the hadronic tensor reads,
\begin{eqnarray}
W^{A\mu\nu}_{nn} \!\!&=&\!\! (-g_\perp^{\mu\nu}) A C_p^A \sum_q Q_q^4 \frac{\alpha_e}{2\pi}
\int dy P(y) \int \frac{d^2l_\perp}{\pi l_\perp^2}  \int{d^3l_q} \int dy_0^- e^{-i(x_B+x_L)p^+y_0^-}  \langle p | \bar{\psi}(y_0^-) \frac{\gamma^+}{2} \psi(0) |p\rangle
\nonumber\\\!\!&\times&\!\! \frac{1}{n!} \left(\prod_{i=1}^{n} \int dY_i^-  \left[ - D_{L1}\frac{\partial}{\partial p_i^-} + \frac{1}{2} D_{L2} \frac{\partial^2}{\partial^2 p_i^-} + \frac{1}{2} D_{T2} {\nabla p_{i \perp}^2}  \right] \right)
 H|_{\mathbf{p}_1 \cdots \mathbf{p}_n = 0}.
\end{eqnarray}
In writing the above expression, we have considered the symmetry of the system and only kept these terms with non-vanishing coefficients.
The transport coefficients in the above equation are defined as,
\begin{eqnarray}
D_{L1} \!\!&=&\!\! g^2 \frac{C_F}{N_c^2 - 1} \int dy^- \frac{\rho}{2p^+} \langle p| i\partial^- A^+(y^-) A^+(0)|p \rangle, \\
D_{L2} \!\!&=&\!\! g^2 \frac{C_F}{N_c^2 - 1} \int dy^- \frac{\rho}{2p^+} \langle p| \partial^- A^+(y^-) \partial^- A^+(0)|p \rangle, \\
D_{T2} \!\!&=&\!\! g^2 \frac{C_F}{N_c^2 - 1} \int dy^- \frac{\rho}{2p^+} \langle p| \partial_\perp A^+(y^-) \partial_\perp A^+(0)|p \rangle. \ \
\end{eqnarray}
These coefficients $D_{L1}$, $D_{L2}$ and $D_{T2}$ are (up to an overall factor) elastic energy loss rate $\hat{e}$, the diffusions of longitudinal and transverse momenta $\hat{e}_2$ and $\hat{q}$ \cite{Qin:2012fua}.
It should be noted that in the above definitions of jet transport coefficients are not gauge-invariant.
The gauge invariant definitions of these coefficients may be obtained by resumming over an arbitrary number of soft gluon insertions.
This will give the Wilson lines between the the gluon field strength operators, which renders the operator product gauge invariant.
More discussion about gauge-invariant definition of jet broadening coefficient $\hat{q}$ can be found in Ref. \cite{Liang:2008rf, Benzke:2012sz}.
We note that the first calculation of the jet transport coefficient $\hat{q}$ within the framework of finite temperature lattice gauge theory was carried out in Ref. \cite{Majumder:2012sh}, whose result is consistent with the values obtained from phenomenological studies of jet quenching at RHIC and the LHC \cite{Burke:2013yra}.

With above expansion, we may now perform the resummation of the number of multiple scatterings,
\begin{eqnarray}
W^{A\mu\nu} = (-g_\perp^{\mu\nu}) A C_p^A \sum_q Q_q^4 \frac{\alpha_e}{2\pi}
\int dy P(y) \int \frac{d^2 l_\perp}{\pi l_\perp^2} \int{d^3l_q} \int dy_0^- e^{-i(x_B+x_L)p^+y_0^-}  \langle p | \bar{\psi}(y_0^-) \frac{\gamma^+}{2} \psi(0) |p\rangle
\phi(L^-, l_q^-, \mathbf{l}_{q\perp}), \ \
\end{eqnarray}
where we have defined the distribution function $\phi(L^-, l_q^-, \mathbf{l}_{q\perp})$,
\begin{eqnarray}
\phi = \sum_{n=0}^{\infty} \frac{1}{n!} \left(\prod_{i=1}^{n} \int dY_i^- \left[ - D_{L1}\frac{\partial}{\partial p_i^-} + \frac{1}{2} D_{L2} \frac{\partial^2}{\partial^2 p_i^-}  + \frac{1}{2} D_{T2} {\nabla p_{i \perp}^2}  \right] \right) H|_{\mathbf{p}_1 \cdots \mathbf{p}_n = 0}. \ \
\end{eqnarray}
We may write down the double differential hadronic tensor as:
\begin{eqnarray}
\frac{dW^{A \mu\nu}}{dy d^2 l_\perp dl_q^- d^2l_{q\perp}} \!\!&=&\!\!  (-g_\perp^{\mu\nu}) A C_p^A \sum_q Q_q^4 \frac{\alpha_e}{2\pi}
\frac{P(y)}{\pi l_\perp^2} f_q(x_B + x_L) \phi(L^-, l_q^-, \mathbf{l}_{q\perp}).
\end{eqnarray}
Note that we may split the hard part into three parts, $H = H_0 + H_- + H_\perp$, representing the three terms in $H$.
Accordingly, we may split the above distribution function $\phi$ into three parts, $\phi = \phi_0 + \phi_- + \phi_\perp$.
The contribution from the term $H_0$, after resumming the number of multiple scatterings, reads,
\begin{eqnarray}
\phi_0  = \exp\left( L^- \left[ D_{L1}\frac{\partial}{\partial l_q^-} + \frac{1}{2} D_{L2} \frac{\partial^2}{\partial^2 l_q^-} + \frac{1}{2} D_{T2} {\nabla_{l_{q\perp}}^2}  \right] \right) \delta(l_q^- - q^-(1-y)) \delta^2(\mathbf{l}_{q\perp} + \mathbf{l}_{\perp}),
\end{eqnarray}
where we have converted the derivatives over $p_i$ to the derivatives over $l_q$.
The above distribution function obviously satisfies the following evolution equation,
\begin{eqnarray}
\frac{\partial \phi_0}{\partial L^-} = \left[ D_{L1}\frac{\partial}{\partial l_q^-} + \frac{1}{2} D_{L2} \frac{\partial^2}{\partial^2 l_q^-} + \frac{1}{2} D_{T2} {\nabla_{l_{q\perp}}^2}  \right] \phi_0(L^-, l_q^-, \mathbf{l}_{q\perp}).
\end{eqnarray}
In fact, such evolution equation describes the three-dimensional evolution of the momentum distribution of a propagating parton which only experiences multiple scattering without any radiation.
The three terms in the evolution equation represent the contributions from longitudinal momentum loss and diffusion, and the diffusion of transverse momentum.
For the case of no radiation, the initial condition is $\phi_0(L^- = 0, l_q^-, \mathbf{l}_{q\perp}) =  \delta(l_q^- - q^-) \delta^2(\mathbf{l}_{q\perp} + \mathbf{l}_{\perp})$, we will return to the result found in Ref. \cite{Qin:2012fua}.
For the case of photon bremsstrahlung, the initial condition changes to $\phi_0(L^- = 0, l_q^-, \mathbf{l}_{q\perp}) = \delta(l_q^- - q^-(1-y)) \delta^2(\mathbf{l}_{q\perp} + \mathbf{l}_{\perp})$. This represents the case that the photon is immediately radiated after the initial hard collisions. For such initial condition, the distribution $\phi_0$ has the following solution,
\begin{eqnarray}
\phi_0 = \frac{e^{-{(l_q^- - q^-(1-y) + D_{L1}L^-)^2}/({2D_{L2}L^-})}}{\sqrt{2\pi D_{L2} L^-}}   \frac{e^{-{(\mathbf{l}_{q\perp} + \mathbf{l}_{\perp})^2}/({2D_{T2}L^-})}}{2\pi D_{T2} L^-}.
\end{eqnarray}
The above solution represents the case that the photon is radiated immediately after the hard collisions, and after that the propagating parton experiences multiple scatterings in the medium.

Now we look at the ($-$)-part of the distribution function $\phi_-(L^-, l_q^-, \mathbf{l}_{q\perp})$ which reads,
\begin{eqnarray}
\phi_- \!\!&=&\!\!  \frac{y(1-y)}{1+(1-y)^2} \sum_{n=1}^{\infty}  \frac{1}{n!}  \sum_{k=1}^n \int dY_k^- F_k \frac{1}{q^-} \left(-D_{L1} - D_{L2} \frac{\partial}{\partial l_q^-} \right)
\nonumber\\ \!\!&\times&\!\!
\left( \prod_{i=1, i\ne k}^n \int dY_i^- \left(D_{L1} \frac{\partial}{\partial l_q^-} + \frac{1}{2} D_{L2} \frac{\partial^2}{\partial^2 l_q^-} + \frac{1}{2} D_{T2} {\nabla_{l_{q\perp}}^2}   \right) \right)
\delta(l_q^- - q^-(1-y)) \delta^2(\mathbf{l}_{q\perp} + \mathbf{l}_{\perp}).
\end{eqnarray}
Assuming the transport coefficients $D_{L1}$, $D_{L2}$ and $D_{T2}$ are position independent, we may perform the $Y_k^-$ integral for the function $F_k$,
\begin{eqnarray}
\int_0^{L^-} dY_k^- F_k \approx \int_0^L dY_k^- 2 \cos(x_L p^+ Y_k^-) = \frac{2\sin(x_L p^+ L^-)}{x_L p^+}.
\end{eqnarray}
The sum over $k$ just give a factor of $n$, which can be used to shift the sum over $n$ from $0$. Thus, the distribution function $\phi_-(L^-, l_q^-, \mathbf{l}_{q\perp})$ is directly related to $\phi_0$ as,
\begin{eqnarray}
\phi_- = \frac{y(1-y)}{1+(1-y)^2} \frac{2\sin(x_L p^+ L^-)}{x_L p^+q^-}  \left(-D_{L1} - D_{L2} \frac{\partial}{\partial l_q^-} \right) \phi_0.
\end{eqnarray}
The analysis of $\phi_\perp(L^-, l_q^-, \mathbf{l}_{q\perp})$ is completely analogous, and we obtain
\begin{eqnarray}
\phi_\perp = y \frac{2\sin(x_L p^+ L^-)}{x_L p^+}  \left(- D_{T2} \frac{\mathbf{l}_\perp \cdot \nabla_{l_{q\perp}}}{l_\perp^2} \right)\phi_0.
\end{eqnarray}
Putting three contributions together, $\phi = \phi_0 + \phi_- + \phi_\perp$, we obtain
\begin{eqnarray}
\phi = \left\{ 1 + \frac{2\sin(x_Lp^+L^-)}{x_Lp^+} \left[\frac{y(1-y)}{1+(1-y)^2}  \left(-\frac{D_{L1}}{q^-} - \frac{D_{L2}}{q^-} \frac{\partial}{\partial l_q^-} \right) + y\left(- \frac{D_{T2}  \mathbf{l}_\perp \cdot \nabla_{l_{q\perp}}}{l_\perp^2} \right) \right] \right\} \phi_0(L^-, l_q^-, \mathbf{l}_{q\perp})
\end{eqnarray}
The double differential hadronic tensor now reads:
\begin{eqnarray}
\label{jet_photon}
\frac{dW^{A \mu\nu}}{dy d^2 l_\perp dl_q^- d^2l_{q\perp}} \!\!&=&\!\!  (-g_\perp^{\mu\nu}) A C_p^A \sum_q Q_q^4 \frac{\alpha_e}{2\pi}
\frac{P(y)}{\pi l_\perp^2} f_q(x_B + x_L) \\\!\!&\times&\!\!
\left\{ 1 + \frac{2\sin(x_Lp^+L^-)}{x_Lp^+} \left[\frac{y(1-y)}{1+(1-y)^2}  \left(-\frac{D_{L1}}{q^-} - \frac{D_{L2}}{q^-} \frac{\partial}{\partial l_q^-} \right) + y\left(- \frac{D_{T2}  \mathbf{l}_\perp \cdot \nabla_{l_{q\perp}}}{l_\perp^2} \right) \right] \right\} \phi_0(L^-, l_q^-, \mathbf{l}_{q\perp}). \nonumber
\end{eqnarray}
\end{widetext}
The above equation represents the main result of this work. The differential spectrum of photons radiated from a hard parton which experiences multiple scatterings in the medium has been expressed as factorized product of the photon splitting function, the three-dimensional distribution of the propagating parton, and a multiplicative factor (in the curly bracket).
The first constant term represents the case that the incoming parton is virtual and radiates photon immediately, and the subsequent multiple scattering experienced by the hard partons is described by $\phi_0$.
The other terms represent the additional contribution in which the propagating parton experiences multiple scatterings and exchanges longitudinal momentum and/or transverse momentum with the medium, and radiates a photon.
Such contribution has been written as the sum of three pieces, corresponding to the contributions from the longitudinal momentum drag, the diffusion of longitudinal momentum and the diffusion of transverse momentum.

One can see that the longitudinal momentum loss (the drag) produces a negative contribution, meaning that the drag tends to suppress the medium-induced photon radiation.
The diffusions of longitudinal and transverse momenta also contribute to the double differential spectrum of the propagating hard quark and the radiated photon.
But we note that in the extremely high energy and collinear limits where the exchanged momentum between the hard parton and medium is small, the integration over $dl_q^-$ from the longitudinal diffusion term and the integration over $d^2l_{q\perp}$ renders zero in the above equation.
This is due to the fact that we have only kept terms up to linear in $K_p^-/q^-$ and $K_{p_\perp}/l_\perp$ in Eq. (\ref{linear_approximation}).
Therefore at this linear order, the longitudinal and transverse diffusion do not contribute to single photon emission spectrum.
However, for the three-dimensional propagation of the hard jet in the dense medium experiencing multiple scatterings and photon bremsstrahlung, Eq. (\ref{jet_photon}) represents the leading contribution.

\section{Conclusions and discussions}

We have studied the photon bremsstrahlung from a hard parton undergoing multiple scatterings when propagating through an extended dense nuclear medium. In particular, we have included the effect of longitudinal momentum exchange from multiple scatterings for the calculation of medium-induced photon emission. We investigated the problem in the framework of deep-inelastic scattering where a virtual photon strikes a hard parton in a nucleon.
The struck parton then propagates through the nucleus, scatters with medium constituents multiple times and radiates a real photon.
The gluon emission from the hard parton is neglected in the current calculation.
In the interest of studying closed form expressions, we have approximated to 
the case of a virtual quark, with radiated photon transverse momentum $l_{\perp} \gg p_{\perp}$, 
the mean momentum exchanged between the quark and the medium per scattering. 
Our result has been factorized into the product of the photon splitting function, the three-dimensional momentum distribution of the hard parton, and a multiplicative factor which encodes the interference effects encountered by the propagating parton through the dense nuclear medium.
This three-dimensional momentum distribution encodes the effect of the longitudinal momentum drag and diffusion, and the transverse momentum diffusion experienced by the propagating hard parton which experiences multiple scatterings with the medium constituents without any radiation.

The present work is the first effort to incorporate the longitudinal momentum exchanges, besides the transverse momentum exchange, in the calculation of medium-induced radiation.
The study of photon emission in this work serves as an intermediate step for the investigation of the same process with medium-induced gluon emission.
While photon does not interact with the medium, its formation incorporates many interesting physics issues such as the LPM interference effect, which will appear in the future calculation of medium-induced gluon radiation from a hard parton in the dense medium.
The medium-induced photon radiation process explored here is also relevant to the study of the photon production and photon-hadron correlations in heavy-ion collisions.
Such phenomenological application is left for future effort.

\section*{Acknowledgments}

This work is supported in part by Natural Science Foundation of China (NSFC) under grant No. 11375072 and in part by the US National Science Foundation under grant No.
PHY-1207918.

%%%%%%%%%%%%%%%%%%%%%%%%%%%%%%%%%%%%%%%%%%%%%%%%%%%%%%%%%%%%%%%%%%%%%%%%%%%%%%%%
%%%%%%%%%%%%%%%%%%%%%%%%%%%%%%%%%%%%%%%%

%%%%%%%%%%%%%%%%%%%%%%%%%%%%%%%%%%%%%%%%%%%%%%%%%%%%%%%%%%%%%%%%%%%%%%%%%%%%%%%%
%%%%%%%%%%%%%%%%%%%%%%%%%%%%%%%%%%%%%%%%%
%\bibHeading{References}
%\bibliographystyle{plain}
\bibliographystyle{h-physrev5}
\bibliography{GYQ_refs}
%%%%%%%%%%%%%%%%%%%%%%%%%%%%%%%%%%%%%%%%%%%%%%%%%%%%%%%%%%%%%%%%%%%%%%%%%%%%%%%%
%%%%%%%%%%%%%%%%%%%%%%%%%%%%%%%%%%%%%%%%
\end{document}